\documentclass[epj]{svjour}
\usepackage{graphicx,amsmath,amssymb,mathptm,multirow,longtable,array}
\newcommand{\knn}{\bar k_{\rm nn} (k)}

\newcommand{\be}{\begin{equation}}
\newcommand{\ee}{\end{equation}}
\begin{document}

\title{Intrinsic degree-correlations in the static model of scale-free networks}
\author{J.-S. Lee \and
K.-I. Goh\and B. Kahng\thanks{\email{kahng@phya.snu.ac.kr}} \and
D. Kim} \institute{School of Physics and Center for Theoretical
Physics, Seoul National University, Seoul 151-747, Korea \\fax :
+82-2-884-3002\\phone : +82-2-880-1326 }
\date{\today}

\abstract{ We calculate the mean neighboring degree function $\bar
k_{\rm{nn}}(k)$ and the mean clustering function $C(k)$ of
vertices with degree $k$ as a function of $k$ in finite scale-free
random networks through the static model. While both are
independent of $k$ when the degree exponent $\gamma \geq 3$, they
show the crossover behavior for $2 < \gamma < 3$ from
$k$-independent behavior for small $k$ to $k$-dependent behavior
for large $k$. The $k$-dependent behavior is analytically derived.
Such a behavior arises from the prevention of self-loops and
multiple edges between each pair of vertices. The analytic results
are confirmed by numerical simulations. We also compare our
results with those obtained from a growing network model, finding
that they behave differently from each other.
 \PACS{ {89.75.Da}{Systems obeying scaling laws} \and {89.75.Fb}{Structure and
organization in complex systems } \and {05.65.+b}{Self-organized
systems}  } }
\maketitle
\section{Introduction}
Structural properties of complex networks have drawn much
attentions recently~\cite{rmp,porto,siam}. Degree, the number of
edges connected to a given vertex, is a primary quantity to
characterize the network structure. In many real-world networks,
degrees are inhomogeneous and their distribution follows a power
law $P_d(k)\sim k^{-\gamma}$. Such networks are called scale-free
(SF) networks~\cite{Barabasi99}. The degree-degree correlation is
also important to characterize network structure. The correlation
between two degrees of vertices connected via an edge is measured
by the mean neighboring degree function $\knn$, which is defined
as the mean degree of neigboring vertices of vertices with degree
$k$~\cite{vespignani}. The correlation among three vertices
centered at a vertex $i$ is measured through the local clustering
coefficient $C_i$, defined as $C_i=2e_i/k_i (k_i-1)$, where $e_i$
is the number of connections among the $k_i$ neighbors of a vertex
$i$. $k_i$ is the degree of the vertex $i$. The clustering
function $C(k)$ is the averaged one of $C_i$ over the vertices
with degree $k$~\cite{Ravasz02a,Vazquez02}.

While Barab\'asi and Albert (BA) introduced a model to generate SF
networks, the model is applied to growing systems where the number
of vertices increases with time~\cite{Barabasi99}. As an extension
of the Erd\H{o}s-R\'enyi (ER) model of random graph to SF
networks~\cite{Erdos59}, where the number of vertices in the
system is fixed, Goh {\it et al.} introduced the so-called static
model~\cite{Goh01}. The term `static' originates from the fact
that the number of vertices $N$ is fixed. The static model was
followed by other similar-type models such as the hidden variable
model ~\cite{rome,Soderberg02,Boguna03}. In the static model, each
vertex $i(i=1,\cdots,N)$ has a prescribed weight $p_i$ summed to
1, which is given as
\begin{equation}
p_i=\frac{i^{-\mu}}{\sum_{j=1}^{N}j^{-\mu}}\approx \frac{1-\mu}{N^{1-\mu}}
i^{-\mu},
\end{equation}
where the Zipf exponent $\mu$ is in the range $0 < \mu < 1$. To
construct the network, in each time step, two vertices $i$ and $j$
are selected with the probability $p_i$ and $p_j$, respectively.
If $i=j$ or an edge connecting $i$ and $j$ already exists, do
nothing, implying that self-loops and multiple edges are not
allowed, respectively. This condition is called the fermionic
constraint hereafter. Otherwise, an edge is added between $i$ and
$j$. This process is repeated $NK$ times. The resulting network is
a scale free one with the degree exponent given as
~\cite{Goh01,nucl}
\begin{equation}
\gamma=1+\frac{1}{\mu}~~. \label{gamma}
\end{equation}
 Since a pair of vertices is selected with rate
$2p_ip_j$, where the factor 2 comes from the two cases of $(i,j)$
and $(j,i)$, one may think that there is no degree correlation,
which is the case we can observe when $\gamma
> 3$. However, when $2 < \gamma < 3$, due to the fermionic
constraint, the degree-degree correlation arises intrinsically. In
this case, the degree correlation occurs for vertices with large
degree, while it is still absent for vertices with small degree.
In this paper, we investigate such degree correlations in terms of
the functions $\knn$ and $C(k)$ in the static model and their
crossover behavior in terms of system size $N$.

Many SF networks in the real-world and artificial networks include
degree correlations within them. For example, the mean neighboring
degree function $\knn$ behaves $\sim k^{-\nu}$ with $\nu
> 0$ for the Internet~\cite{vespignani} and the protein interaction network~\cite{Maslov02}, while
$\nu < 0$ for social networks such as the coauthorship network.
The case with $\nu > 0$ ($\nu < 0$) is called disassortative
(assortative) mixing~\cite{Newman02}. When a network contains
hierarchical and modular structure within it, it is suggested that
the mean clustering function $C(k)$ behaves as $C(k)\sim
k^{-\beta}$ for large $k$ as observed in metabolic networks and
the hierarchy model~\cite{Ravasz02a,Rowasz02}. Occurrence of such
degree correlations in real-world networks may be related to their
own functional details. For example, the assortativity of the
social network arises from the social relationship between bosses,
while the disassortativity of the Internet comes from the network
design to allow data packets flow efficiently. The three-degree
correlation may be related to the control system in biological
network such as the feed-back or feed-forward loop
structure~\cite{Milo02}. Such degree correlations in real world
networks appear in the combination of those due to the fermionic
constraint and their functionalities. On the other hand, the
static model is frequently used to study various dynamical
properties of complex networks. Therefore, the knowledge of the
intrinsic degree correlations we study here would be helpful in
understanding the degree-correlation a SF network has for
functional activity. For the purpose, Catanzaro and
Pastor-Satorras~\cite{mich} studied the degree-correlation
function $\knn$ for the static model, but their study relies on
numerics in the end. Here we present analytic solutions for $\knn$
as well as clustering function $C(k)$. We mention that $\bar
k_{\rm{nn}}(k)$ for a related model was analyzed by Park and
Newman~\cite{Park03} while $\bar k_{\rm{nn}}(k)$ and $C(k)$ for
the BA-type growing network models are studied by Barrat and
Pastor-Satorras~\cite{barrat} using the rate equation
approach~\cite{szabo}. On the other hand, it was desirable to
introduce uncorrelated SF network as a null model to check the
correctness of analytic solutions in various problems on SF
networks. For the purpose, Bogu\~n\'a {\it et al.}~\cite{boguna}
and Catanzaro {\it et al.}~\cite{uncorr} introduced a way to
construct uncorrelated SF network by restricting degree of each
vertex to be less than the cutoff value $k_c$, beyond which the
intrinsic correlation arises in $2 < \gamma < 3$. The cutoff value
they used scales as $\sim N^{1/2}$, independent of $\gamma$, which
was based on the configuration model introduced by Molloy and
Reed~\cite{MR}. Such cutoff is also implicit in the model
introduced by Chung and Lu~\cite{Ching}. However, we show that
while the natural cutoff of the static model is $\sim
N^{1/(\gamma-1)}$, the vertex correlations appear for degrees
larger than a crossover value, $k_{c_1} \sim
N^{(\gamma-2)/(\gamma-1)}$, which is smaller than $N^{1/2}$ for
$2<\gamma<3$. We mention that $k_{c_1}\sim \rm{const}$($\sim
N^{1/2}$) as $\gamma \rightarrow 2$($\gamma \rightarrow 3$) so
that for $\gamma \rightarrow 2$ all the nodes have nontrivial
vertex correlations and for $\gamma \rightarrow 3$, there are no
correlations.

In Section 2, we derive the mean neighboring degree function
$\knn$ and the mean clustering function $C(k)$ analytically.
Comparisons between the results of our analytic derivations and
numerical simulations are given in Section 3. Section 4 summarizes
our results.

\section{Analytic Results}
\begin{figure}
\resizebox{1\columnwidth}{!}{\includegraphics{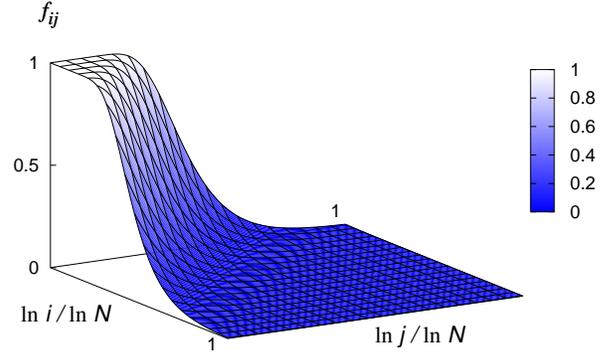} }
\caption{The connection probability $f_{ij}$ of an edge has two
distinct regions where $f_{ij}\approx 1$ or $\approx 2KNp_ip_j$
due to the fermionic constraint when $2<\gamma <3$.}
\label{fig:f_ij}       
\end{figure}

In the static model, the notion of the grand canonical ensemble is
applied~\cite{nucl}, where the number of edges is a fluctuating
variable while keeping the SF nature of the degree distribution.
Each pair of vertices $(i,j)$ is connected independently with the
probability $f_{ij}$, given by
\begin{equation}
f_{ij}=1-e^{-2NKp_{i}p_{j}}, \label{f-ij}
\end{equation}
because the probability that vertices $i$ and $j$ $(i\neq j)$ are
not connected after $NK$ trials is given by
$(1-2p_ip_j)^{NK}\simeq e^{-2NKp_ip_j}$. That is, if we denote the
adjacency matrix element by $a_{ij}$ $(=0,1)$ then its ensemble
average is given by $f_{ij}$; i.e., $\langle a_{ij}
\rangle=f_{ij}$, $\langle \cdots \rangle$ denoting the grand
canonical ensemble average. For $i=j$, $f_{ij}=0$ because of the
prevention of self-loops. Since $2NKp_ip_j\sim K
N^{2\mu-1}/(ij)^{\mu}$ for finite $K$, when $0 < \mu < 1/2$,
corresponding to the case $\gamma > 3$, $2KNp_ip_j$ is small in
the thermodynamic limit, therefore, \be f_{ij}\approx 2KNp_ip_j.
\label{boson} \ee This is called the bosonic limit. On the other
hand, when $1/2 <\mu <1$, corresponding to the case $2 <\gamma
<3$, $2KNp_ip_j$ may diverge in the thermodynamic limit,
therefore, $f_{ij}$ is not necessarily of the form of
Eq.~(\ref{boson}), but it reduces to \be f_{ij}\approx\left\{
\begin{tabular}{ll}
$1$ & \quad when \quad $ij\ll N^{2-\frac{1}{\mu}}$,\\
$2KNp_ip_j$ & \quad when \quad $ij\gg N^{2-\frac{1}{\mu}}$.
\end{tabular}
\right. \ee This is the manifestation of the fermionic constraint,
the prevention of multiple edges. Thus, for $2<\gamma <3$,
$f_{ij}$ has two distinct regions in the ($i$,$j$) plane as shown
in Fig.~\ref{fig:f_ij}.

\subsection{Degree and degree distribution:}
The degree $k_{i}$ of a vertex $i$ is given in terms of the
adjacency matrix as $k_{i} = \sum_{j} a_{ij}$. For completeness,
we present here known results for the mean degree $\langle
k_i\rangle$~\cite{nucl}. It is obtained through the formula
$\langle k_i \rangle=\sum_{j\neq i}f_{ij}$ which can be evaluated
by using its integral form as
\begin{equation}
\langle k_i \rangle=\sum_{j\ne i}f_{ij} \approx \int_{1}^{N}dj
f_{ij} = \frac{1}{\mu} a^{\frac{1}{\mu}} N^{1-\frac{1}{2\mu}}
    \int_{a N^{-\frac{1}{2}}}^{aN^{\mu-\frac{1}{2}}}dy
    \frac{1-e^{-xy}}{y^{\gamma}},
    \label{eqq2}
\end{equation}
where $\gamma=1+\frac{1}{\mu}$, $x=a N^{\mu-\frac{1}{2}}i^{-\mu}$
and $y=a N^{\mu-\frac{1}{2}}j^{-\mu}$ with
\begin{equation}
    a=\sqrt{2K(1-\mu)^2}.
    \label{eqq-a}
\end{equation}
The integral in Eq.(\ref{eqq2}) denoted as $I(x)$ is evaluated as
\begin{eqnarray}
    I(x)&=&\int_{a N^{-\frac{1}{2}}}^{a N^{\mu-\frac{1}{2}}}dy
    \frac{1-e^{-xy}}{y^{\gamma}}\nonumber \\ &\approx& \left\{
    \begin{array}{ll} \frac{a^{2-\gamma}
        N^{\frac{\gamma-2}{2}}}{\gamma-2}(1-N^{\mu-1})x &~~~ {\rm
        for}~~~x<
        {1}/{a N^{\mu-\frac{1}{2}}}, \\
        \frac{a^{2-\gamma} N^{\frac{\gamma-2}{2}}}{\gamma-2}x +q_0(\gamma)
        x^{\gamma-1} &~~~ {\rm for}~~~ x>{1}/{a N^{\mu-\frac{1}{2}}},
    \end{array} \right.
    \label{I-y}
\end{eqnarray}
with $q_0(\gamma)\equiv
\int_{0}^{\infty}dr{(1-e^{-r}-r)}/{r^{\gamma}}$, which is a
negative constant.
Therefore, we obtain that \be \langle k_i \rangle \approx
2K(1-\mu)\Big(\frac{N}{i}\Big)^{\mu}+{\mathcal {A}},
\label{eq:k_i} \ee where $\mathcal A$ is the correction, of which
the leading term is
\begin{equation}
    {\mathcal A} \approx \left\{ \begin{array}{ll}
    -2K(1-\mu)N^{2\mu-1}/i^{\mu}
    &~~~{\rm for}~~~i> a^{\frac{2}{\mu}}N^{2-\frac{1}{\mu}}, \\
    \big(2K(1-\mu)^2\big)^{\frac{1}{\mu}}
    N^{2-\frac{1}{\mu}}q_0(\gamma)/(i \mu) &~~~{\rm for}~~~
    i<a^{\frac{2}{\mu}}N^{2-\frac{1}{\mu}}.
    \end{array} \right.
    \label{eqq9}
\end{equation}
This is negligible compared with the first term on the right hand
side of Eq.~(\ref{eq:k_i}) in the thermodynamic limit, $N
\rightarrow \infty$. The average degree is then \be \bar k \equiv
\frac{2 \langle L \rangle}{N}=\frac{1}{N}\sum_i \langle k_i
\rangle=2K, \ee where $\langle L \rangle$ is the mean number of
edges in the grand canonical ensemble. From Eq.(\ref{eq:k_i}), one
can easily obtain that the degree exponent is related to the Zipf
exponent $\mu$ as $\gamma=1+1/\mu$ given in Eq.~(\ref{gamma})
\subsection{Mean neighboring degree function $\knn$:}
We pay attention to the case $2<\gamma<3$. To evaluate the mean
neighboring degree function $\knn$, we first calculate the mean
neighboring degree in terms of $i$, i.e., $\bar k_{\rm nn} (i)$
and convert it to $\knn$ by using the relation of $\langle k_i
\rangle$ versus $i$. To proceed, we use the expression, \be \bar
k_{\rm{nn}}(i)= {\Big \langle} \frac{\sum_{j~\in {\rm ~nn~of~}i}
k_j}{ k_i} {\Big \rangle} = {\Big \langle} \frac{\sum_{j,k}
a_{ij}a_{jk}}{\sum_{j}a_{ij}} {\Big \rangle} \approx \frac{\langle
\sum_{j,k}a_{ij}a_{jk} \rangle}{\langle k_i \rangle},
\label{eqq9.5} \ee where the ensemble average is applied to the
numerator and the denominator separately. Its validity is checked
numerically, which is shown in Section III. The denominator was
already derived, and the numerator is evaluated as follows:
\begin{equation}
{\Big \langle} \sum_{j,k}a_{ij}a_{jk} {\Big \rangle}=
\sum_{j,k\neq i}f_{ij}f_{jk} + \sum_{j}f_{ij} \approx
    \int_{1}^{\infty} dj \int_{1}^{\infty}dk~ f_{ij}f_{jk}+\langle k_{i} \rangle.
    \label{eqq11}
\end{equation}
where $a_{ij}^2 = a_{ij}$ is used and the double sum is
approximated by the double integral. The validity of the
transformation from the discrete double sum to the double
integration is discussed in Appendix A where $2<\gamma<3$. Such an
approximation introduces at most an $O(1)$ factor on the amplitude
of the leading order terms for large $N$, as will be mentioned
below. Applying the similar method used in Eq.~(\ref{eqq2}), we
evaluate the integration in Eq.~(\ref{eqq11}) as
\begin{eqnarray}
    \int_{1}^{\infty}&dj & \int_{1}^{\infty}dk f_{ij}f_{jk} =
    \frac{a^{\frac{2}{\mu}}
    N^{(2-\frac{1}{\mu})}}{\mu^2}\times \nonumber \\
    &\times& \int_{a
    N^{-\frac{1}{2}}}^{a N^{\mu-\frac{1}{2}}} dy
    \frac{1-e^{-xy}}{y^{\gamma}} \int_{a
    N^{-\frac{1}{2}}}^{a N^{\mu-\frac{1}{2}}} dz
    \frac{1-e^{-yz}}{z^{\gamma}},
    \label{eqq12}
\end{eqnarray}
where $x\equiv a N^{\mu-{1}/{2}}i^{-\mu}$ is in the range $a
N^{-{1}/{2}} < x < a N^{\mu-{1}/{2}}$ (see the definition of $a$ in
Eq.(\ref{eqq-a})). The last part of the integral in
Eq.~(\ref{eqq12}) is $I(y)$ defined in Eq.~(\ref{I-y}). Therefore,
we substitute the leading term of Eq.~(\ref{I-y}) into
Eq.~(\ref{eqq12}) and obtain
\begin{equation}
    \int_{1}^{\infty}dj \int_{1}^{\infty}dk f_{ij}f_{jk}
      \approx  \frac{a^{1+\frac{1}{\mu}}
    N^{\frac{3}{2}-\frac{1}{2\mu}}}{\mu(1-\mu)}x^{\gamma-2}
    \int_{a x N^{-\frac{1}{2}}}^{a x
    N^{\mu-\frac{1}{2}}} dq \frac{1-e^{-q}}{q^{\gamma-1}}.
    \label{eqq13}
\end{equation}
in which we change the variable of integration as $q=xy$. The
integral in the right hand side of Eq.~(\ref{eqq13}) is evaluated
in three parts as
\begin{eqnarray}
    \int_{a x N^{-\frac{1}{2}}}^{a x
    N^{\mu-\frac{1}{2}}} &dq& \frac{1-e^{-q}}{q^{\gamma-1}}
    = \int_{0}^{\infty}dq
    \frac{1-e^{-q}}{q^{\gamma-1}}\nonumber \\
    &-&\int_{0}^{a x
   N^{-\frac{1}{2}}}dq \frac{1-e^{-q}}{q^{\gamma-1}} -
    \int_{a x N^{\mu-\frac{1}{2}}}^{\infty}dq
    \frac{1-e^{-q}}{q^{\gamma-1}}.
    \label{eqq13_1}
\end{eqnarray}
The first term is denoted as
\begin{equation}
    q_1(\gamma) \equiv \int_{0}^{\infty} dq \frac{1-e^{-q}}{q^{\gamma-1}},
    \label{eqq14}
\end{equation}
which is a positive constant for $2<\gamma<3$. The second term is,
since $ax N^{-{1}/{2}} \ll 1$,
\begin{equation}
    \int_{0}^{a x N^{-\frac{1}{2}}}dq
    \frac{1-e^{-q}}{q^{\gamma-1}}
    \approx \frac{a^{3-\gamma}N^{-(3-\gamma)/2}}{3-\gamma}
    x^{3-\gamma}.\label{eqq15}
\end{equation}
The last term is calculated as, when $x \ll {1}/{a
N^{\mu-\frac{1}{2}}}$ ($i \gg
a^{\frac{2}{\mu}}N^{2-\frac{1}{\mu}}$),
\begin{eqnarray}
    \int_{a x N^{\mu-\frac{1}{2}}}^{\infty}dq
    \frac{1-e^{-q}}{q^{\gamma-1}} &=&
    \int_{0}^{\infty}dq \frac{1-e^{-q}}{q^{\gamma-1}}-\int_{0}^{x
    a N^{\mu-\frac{1}{2}}}dq \frac{1-e^{-q}}{q^{\gamma-1}}\nonumber \\
    &\approx& q_1(\gamma)-\frac{a^{3-\gamma}
    N^{(3-\gamma)(\mu-\frac{1}{2})}}{3-\gamma}x^{3-\gamma},
\label{xx}
\end{eqnarray}
and, when $x \gg {1}/{a N^{\mu-\frac{1}{2}}}$ ($i \ll
a^{\frac{2}{\mu}}N^{2-\frac{1}{\mu}}$),
\begin{eqnarray}
    \int_{a x N^{\mu-\frac{1}{2}}}^{\infty}dq
    \frac{1-e^{-q}}{q^{\gamma-1}} & \approx &
     \frac{a^{1-\frac{1}{\mu}}N^{(1-\frac{1}{\mu})(\mu-\frac{1}{2})}}
     {\frac{1}{\mu}-1}x^{2-\gamma}.
    \label{eqq16}
\end{eqnarray}
Combining all the contributions, when $x \ll {1}/{a
N^{\mu-\frac{1}{2}}}$, the second term on the right hand side of
Eq.~(\ref{eqq13_1}) becomes the leading order term, while when $x
\gg {1}/{a N^{\mu-\frac{1}{2}}}$, the first one does. Thus,
\begin{equation}
    \int_{1}^{\infty}dj\int_{1}^{\infty}dk~ f_{ij}f_{jk}
    \approx \left\{ \begin{array}{ll}
    \frac{a^4 N^{3\mu-1}}{(1-\mu)(2\mu-1)}i^{-\mu}
    &~~~{\rm for}~~~i> a^{\frac{2}{\mu}}N^{2-\frac{1}{\mu}}, \\
    q_1(\gamma)\frac{a^{\frac{2}{\mu}}N^{3-\mu-\frac{1}{\mu}}}{\mu(1-\mu)}i^{-1+\mu}
    &~~~{\rm for}~~~i<a^{\frac{2}{\mu}}N^{2-\frac{1}{\mu}}.
    \end{array} \right.
    \label{eqq9}
\end{equation}
The second term $\langle k_{i} \rangle$ on the right-hand-side of
Eq.~(\ref{eqq11}) can be neglected compared with Eq.~(\ref{eqq9})
for all range of $2<\gamma<3$ and $i$. Therefore,
\begin{equation}
    {\bar k_{\rm nn}}(i) \approx  \left\{ \begin{array}{ll}
    \frac{a^2}{2\mu-1}N^{2\mu-1} &~~~{\rm when}~~~i >
    a^{\frac{2}{\mu}} N^{2-\frac{1}{\mu}}, \\
    q_1(\gamma) a^{\frac{2}{\mu}-2} N^{3-2\mu-\frac{1}{\mu}} i^{2\mu-1}/\mu
    &~~~{\rm when}~~~ i < a^{\frac{2}{\mu}} N^{2-\frac{1}{\mu}},
    \end{array} \right.
    \label{eqq16-1}
\end{equation}
and using Eq.~(\ref{eq:k_i}) for $k= \langle k_{i} \rangle$,
\begin{equation}
  {\bar k_{\rm nn}}(k) \approx \left\{\begin{array}{ll}
  \frac{a^2}{2\mu-1}N^{2\mu-1} &~~~{\rm when}~~~ k <
  {N^{1-\mu}}, \\
  q_1(\gamma)2K(1-\mu)^{\frac{1}{\mu}}N^{2-\frac{1}{\mu}}k^{-2+\frac{1}{\mu}}/\mu
  &~~~{\rm when}~~~k > N^{1-\mu}. \end{array} \right.
  \label{eqq17}
\end{equation}
Here we note that the coefficient of $N^{2\mu-1}$ when $i >
a^{\frac{2}{\mu}} N^{2-\frac{1}{\mu}}$(or when $k < {N^{1-\mu}}$)
is not exact but is in between ${a^2}/{(2\mu-1)}$ and ${2\mu
a^2}/{(2\mu-1)}$ as explained in Appendix A
  (see Eq.(\ref{appendix5})). In
terms of the degree exponent $\gamma$ we rewrite Eq.~(\ref{eqq17})
as
\begin{eqnarray}
    {\bar k_{\rm nn}} (k) \sim \left\{\begin{array}{ll}
     N^{(3-\gamma)/(\gamma-1)} &~~~{\rm when}~~~ k > k_{c1}, \\
     N^{3-\gamma} k^{-(3-\gamma)}
     &~~~{\rm when}~~~k < k_{c1}, \end{array} \right.
     \label{eqq17-1}
\end{eqnarray}
where the crossover degree $k_{c1}$ scales as $k_{c1} \sim
N^{(\gamma-2)/(\gamma-1)}$.
\subsection{Clustering function $C(k)$: }
The clustering function $C(k)$ is the mean of the
local clustering coefficient $C_i$ over the vertices with degree $k$. To
calculate $C(k)$, we first calculate $C_i$ and
convert it to $C(k)$ by using the relation Eq.~(\ref{eq:k_i}).
As we introduced before, $C_i$ is defined as $C_i=2e_i/k_i(k_i-1)$,
where $e_i$ is the number of connections among the $k_i$ neighbors. In
the grand canonical ensemble, $C_i$ is calculated as
\begin{equation}
C_i={\Big \langle} \frac{e_i}{k_i(k_i-1)/2} {\Big \rangle}.
\label{c_i}
\end{equation}
However, we approximate it as
\begin{equation}
C_i \approx \frac{\langle {e_i} \rangle}{\langle k_i(k_i-1)/2
\rangle}, \label{c_i_approx}
\end{equation}
which enables us to calculate it analytically. The validity of
this approximation is checked numerically in Section III. We
evaluate the denominator and numerator separately.

The denominator is evaluated as
\begin{eqnarray}
\Big\langle \frac{k_i(k_i-1)}{2} &\Big\rangle&=\frac{1}{2} \sum_{j
\neq k(\neq i)}f_{ij}f_{ik}\nonumber \\
&\approx& \frac{1}{2} \sum_{j, k(\neq i)}f_{ij}f_{ik} \approx
2K^2(1-\mu)^2N^{2\mu}/i^{2\mu}. \label{eqq19}
\end{eqnarray}
The numerator is evaluated as
\begin{eqnarray}
    &&\langle e_i \rangle = \frac{1}{2} \sum_{j \neq k (\neq i)}
    f_{ij}f_{jk}f_{ki} ~~~~~~~~~~~~~~~~~~~~~~~~~~~~~~~~~~~~~~~~~~~~~~~~~~~~~~~~\nonumber \\
    &&\approx {1 \over 2}\sum_{j,k} f_{ij}f_{jk}f_{ki} \approx \frac{1}{2}
    \int_{1}^N dj \int_{1}^{N} dk f_{ij}f_{jk}f_{ki}
    = \frac{a^{\frac{2}{\mu}}N^{2-\frac{1}{\mu}}}{2\mu^2}\times
    \nonumber \\
    &&\times \int_{aN^{-\frac{1}{2}}}^{aN^
    {\mu-\frac{1}{2}}} dy \int_{a N^{-\frac{1}{2}}}^{a N^{\mu-\frac{1}{2}}}dz \frac{(1-e^{-xy})
    (1-e^{-yz})(1-e^{-zx})}{y^{\gamma}z^{\gamma}}.
\label{eqq20}
\end{eqnarray}
Possible errors involved in using the integral form for the double
sum is discussed in Appendix A and will be mentioned below. The
evaluation of the integrals of Eq.(\ref{eqq20}) is carried out
depending on the magnitude of $x$. When $x \gg {1}/{a
N^{\mu-\frac{1}{2}}}$, we obtain
\begin{eqnarray}
    &&\int_{a N^{-\frac{1}{2}}}^{a N^{\mu-\frac{1}{2}}} dz
    \frac{(1-e^{-yz})(1-e^{-zx})}{z^{\gamma}} \nonumber \\
    &&\approx \left\{
    \begin{array}{ll} q_0(\gamma) \left( x^{\gamma-1}-(x+y)^{\gamma-1}
    \right) &~~{\rm when}~~ y <  {1}/{a N^{\mu-
    \frac{1}{2}}}, \\ q_0(\gamma) \left( x^{\gamma-1}+y^{\gamma-1}-
    (x+y)^{\gamma-1} \right) &~~{\rm when}~~y > {1}/{a
    N^{\mu-\frac{1}{2}}}.
    \end{array} \right.
    \nonumber \\
    \label{eqq22}
\end{eqnarray}
Thus $\langle e_i \rangle$ is written as
\begin{equation}
    \langle e_i \rangle=-\frac{a^{\frac{2}{\mu}}N^{2-\frac{1}{\mu}}}{2
    \mu^2}q_0(\gamma) ({\mathcal B}+{\mathcal C}),
    \label{eqq23}
\end{equation}
where ${\mathcal B}$ and ${\mathcal C}$ are expressed in the integral forms as
\begin{eqnarray}
    {\mathcal B}&=&\int_{a N^{-\frac{1}{2}}}^{a
    N^{\mu-\frac{1}{2}}} dy
    \frac{1-e^{-xy}}{y^{\gamma}} \left[(x+y)^{\gamma-1}-
    x^{\gamma-1}-y^{\gamma-1} \right] \nonumber \\
    &=&\int_{a N^{-\frac{1}{2}}/x}^{1}dq
    \frac{(\gamma-1)(1-e^{-x^2q})}{q^{\gamma-1}} \nonumber \\
    &-&\int_{a N^{-\frac{1}{2}}/x}^{1}dq \frac{1-e^{-x^2 q}}{q}
    \nonumber \\
    &+&\int_{1}^{a N^{\mu-\frac{1}{2}}/x} dq
    \left( \frac{\gamma-1}{q^2}-\frac{1}{q^{\gamma}}\right)
    (1-e^{-x^2q})\nonumber \\
    \label{eqq24}
\end{eqnarray}
and
\begin{equation}
    {\mathcal C}=-q_0(\gamma)\int_{a N^{-\frac{1}{2}}}^{{1}/{a
    N^{\mu-\frac{1}{2}}}} dy  \frac{1-e^{-xy}}{y^{\gamma}}
    y^{\gamma-1}.
    \label{eqq24-1}
\end{equation}
Even in the region of $x > 1/aN^{\mu-\frac{1}{2}}$, the leading
term is determined depending on the magnitude of $x$. When $x
>1 > 1/aN^{\mu-\frac{1}{2}}$, the first term of the
integral ${\mathcal B}$ is the most dominant one compared with the
other terms of ${\mathcal B}$ and ${\mathcal C}$, which is
evaluated as $\approx (\gamma-1) q_1(\gamma)x^{2(\gamma-2)}$. When
$1 > x > 1/aN^{\mu-\frac{1}{2}}$, however, the third term is most
dominant and evaluated as $\approx 2(\gamma-1)\ln(1/x)x^2$.
Therefore, the numerator is evaluated as
\begin{equation}
    \langle e_i \rangle
\approx \left\{
    \begin{array}{ll}
    \frac{a^{\frac{2}{\mu}} N^{2-\frac{1}{\mu}}}{2 \mu^2}
    [-q_0(\gamma)q_1(\gamma)(\gamma-1)x^{2(\gamma-2)}],
    ~~~~{\rm when}~~ x > 1,
    \\
    \frac{a^{\frac{2}{\mu}} N^{2-\frac{1}{\mu}}}{2 \mu^2}
    2 (\gamma-1)\ln (1/x)x^2, ~~{\rm when}~~1 > x > {1}/{a
    N^{\mu-\frac{1}{2}}}.
    \end{array} \right.
    \label{eqq30}
\end{equation}
Thus, we get $C_i$ in the region of $x > 1/aN^{\mu-\frac{1}{2}}$
to the leading order as
\begin{eqnarray}
    C_i& \approx & \left\{ \begin{array}{ll}
    N^{1-\frac{1}{\mu}}\ln ({i^{\mu}}/{N^{\mu-\frac{1}{2}}}),
    &~{\rm when~~} N^{1-\frac{1}{2\mu}}< i < N^{2-\frac{1}{\mu}}, \\
    N^{5-4\mu-\frac{2}{\mu}} i^{2(2\mu-1)}, &
    ~{\rm when~~} i < N^{1-\frac{1}{2\mu}}.
    \end{array} \right. \label{c-ii}
    \end{eqnarray}
    Equivalently,
    \begin{eqnarray}
    C(k)& \approx & \left\{\begin{array}{ll}
    N^{1-\frac{1}{\mu}}\ln (N^{\frac{1}{2}}/{k}), &~{\rm when~~} N^{1-\mu}< k< N^{\frac{1}{2}}, \\
    N^{3-\frac{2}{\mu}} k^{-2(3-\gamma)}, &~{\rm when~~} N^{\frac{1}{2}}<k.
    \end{array} \right.
    \label{c-k}
\end{eqnarray}
Let us consider the case of $x \ll {1}/{a N^{\mu-\frac{1}{2}}}$
($i \gg N^{2-\frac{1}{\mu}}$). In this case,
\begin{equation}
    \int_{a N^{-\frac{1}{2}}}^{a N^{\mu-\frac{1}{2}}} dz
    \frac{(1-e^{-yz})(1-e^{-zx})}{z^{\gamma}} \approx
    -q_0(\gamma)\big((x+y)^{\gamma-1}-y^{\gamma-1}\big),
\end{equation}
for $y>{1}/{a N^{\mu-\frac{1}{2}}}$ and is almost negligible for
$y < {1}/{a N^{\mu-\frac{1}{2}}}$. Thus
\begin{eqnarray}
    \langle e_i \rangle &=& -\frac{q_0(\gamma)a^{\frac{2}{\mu}}N^{2-\frac{1}{\mu}}}{2 \mu^2}
    \int_{{1}/{a N^{\mu-\frac{1}{2}}}}^{a N^{\mu-\frac{1}{2}}}
    [(x+y)^{\gamma-1}-y^{\gamma-1}]
    \frac{1-e^{-xy}}{y^{\gamma}} \nonumber \\
    & \approx & -\frac{q_0(\gamma)a^{\frac{1}{\mu}(\gamma-1)} N^{2-\frac{1}{\mu}}}{2\mu^2}x^2 \ln(a^2
    N^{2\mu-1}).
    \label{eqq34}
\end{eqnarray}
Therefore we get when $x \ll 1/{a N^{\mu-\frac{1}{2}}}$ (i.e., $i
\gg a^{\frac{2}{\mu}}N^{2-\frac{1}{\mu}}$, or $k \ll N^{1-\mu}$),
\begin{equation}
    C_i = C(k) \sim A N^{1-\frac{1}{\mu}}\ln N,
    \label{eqq35}
\end{equation}
with $A=-q_0(\gamma) a^{\frac{2}{\mu}-2} (\gamma-2)^2 (3-\gamma)$,
when $k<N^{1-\mu}$. In Appendix A, the error introduced in
Eq.~(\ref{eqq20}) is estimated and is found not to change
Eq.~(\ref{c-k}). For Eq.~(\ref{eqq35}), however, we find $C(k)
\sim AN^{1-\frac{1}{\mu}} ( \ln N +D)$ with an undetermined
constant $D$. Eqs.~(\ref{c-k}) and ~(\ref{eqq35}) are the main
results of this subsection.

\section{Numerical simulations}
We now discuss numerical check of the analytic results derived in
Section 2. For the purpose, the static model network is generated
with $K=2$ and $\mu=2/3$ $(\gamma=2.5)$ and with varying system
size $N$. All data below are averaged over $10^4$ network
configurations.
\begin{figure}
\resizebox{0.9\columnwidth}{!}{\includegraphics{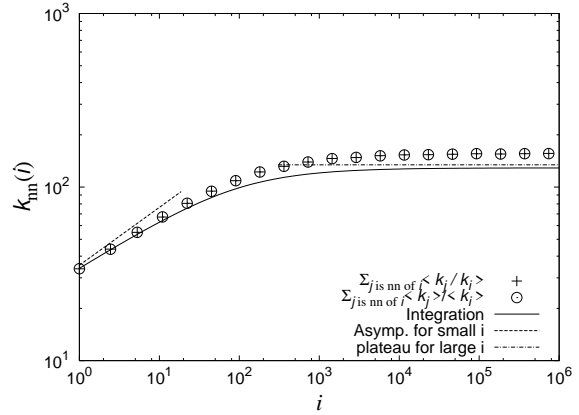}}
\caption{Plot of ${\bar k_{\rm nn}}(i)$ versus $i$. To check the
validity of the approximation Eq.~(\ref{eqq9.5}), we plot
$\sum_{j~\in {\rm {~nn~of~}} i} {\Big \langle} \frac{k_j}{k_i}
{\Big \rangle}$ ($+$) and $ \frac{\sum_{j~\in {\rm ~nn~of~}~i}
\langle k_j \rangle} {\langle k_i \rangle}$=
$\frac{\sum_{j,k}f_{ij}f_{jk}}{\sum_{j}f_{ij}}+1$ ($\circ$) for
$N=10^6$. We can see that the approximation is valid. We compare
them with $\frac{\int \int djdk f_{ij}f_{jk}}{\int dj f_{ij}}+1$
(solid line). They agree with each other for small $i$, however,
it is in disagreement in the plateau region as expected in
Appendix A. We also plot with the first leading term presented in
the text in the asymptotic regions with dot-dashed line and dashed
line, respectively.} \label{fig1}
\end{figure}
\begin{figure}
\resizebox{0.9\columnwidth}{!}{\includegraphics{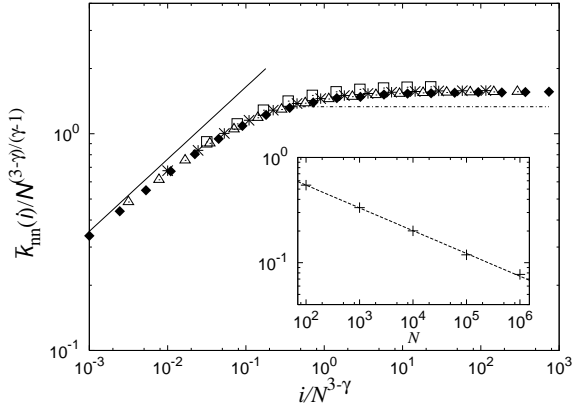}}
\caption{Plot of size-dependent behavior of ${\bar k_{\rm
nn}}(i)$. Data of different network sizes $N=10^3(\square)$,
$10^4(\ast)$, $10^5(\vartriangle)$ and $10^6(\blacklozenge)$
collapse into a single curve in the scaling plot. Inset: Plot of
the difference between the leading order analytic expression and
the simulation value of $i=1$, divided by the simulation value. As
$N$ increases, the relative difference decreases, showing that the
analytic solution converges to the numerical result.} \label{fig2}
\end{figure}
\begin{figure}
\resizebox{0.9\columnwidth}{!}{\includegraphics{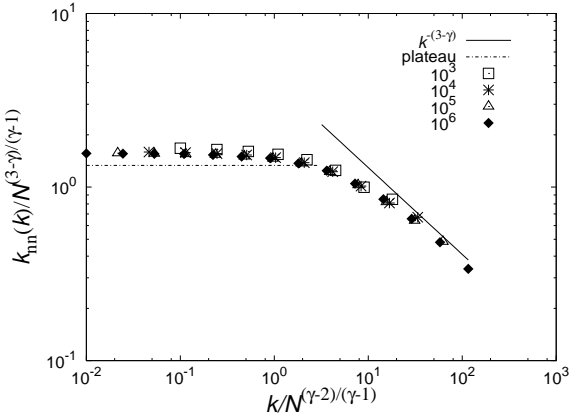}}
\caption{Plot of ${\bar k_{\rm nn}}(k)$ versus $k$ for different
system sizes $N=10^3(\square)$, $10^4(\ast)$, $10^5(\vartriangle)$
and $10^6(\blacklozenge)$. Data for different system sizes
collapse in the scaling plot. Solid and dot-dashed lines indicate
the analytic results of leading order for large and small $k$,
respectively.} \label{fig3}
\end{figure}
For the case of $\knn$, we first check the approximation,
Eq.~(\ref{eqq9.5}). To proceed, we measure $\sum_{j \in ~{\rm
nn~of~}i} \langle k_j /k_i \rangle$ and $\sum_{j \in ~{\rm
nn~of~}i} \langle k_j \rangle / \langle k_i \rangle$ separately in
Fig.\ref{fig1}, finding that the data overlap and the
approximation is valid. Next we directly enumerate the function,
$\int_{1}^{N}dj \int_{1}^{N}dk f_{ij}f_{jk} / \int_{1}^{N}dj
f_{ij}+1$ (solid line) and compare it with the evaluation (dashed
line) within leading order, Eq.~(\ref{eqq17}). The extra term of
`1' comes from the 2nd term $\langle k_i \rangle$ of
Eq.~(\ref{eqq11}). For small $i$, the two lines seem to be
consistent, however, for large $i$, they somewhat deviate in the
intermediate region of $i$. However, we confirm that our analytic
solution is valid within leading order by the finite size scaling
plot. In Fig.\ref{fig2}, we plot $\bar k_{\rm{nn}}(i)$ for
different $N=10^3, 10^4$, $10^5$ and $10^6$ finding that the data
collapse into a single curve by the rescalings of $i \rightarrow
i/N^{2-{1}/{\mu}}$ and $\bar k_{\textrm{nn}}(i) \rightarrow {\bar
k_{\rm{nn}}}(i)/N^{2\mu-1}$. Moreover, we find as we increase the
system size that the numerical simulation data approaches our
analytic solution for small $i$(inset of Fig.\ref{fig2}). We also
check the behavior of $\knn$ numerically. Under the rescaling of
$k \rightarrow k/N^{1-\mu}$ and $\bar k_{\rm{nn}}(k) \rightarrow
\bar k_{\rm{nn}}(k)/N^{2\mu-1}$, the data for different system
sizes collapse well, confirming the validity of our analytic
result.
\begin{figure}
\resizebox{0.9\columnwidth}{!}{\includegraphics{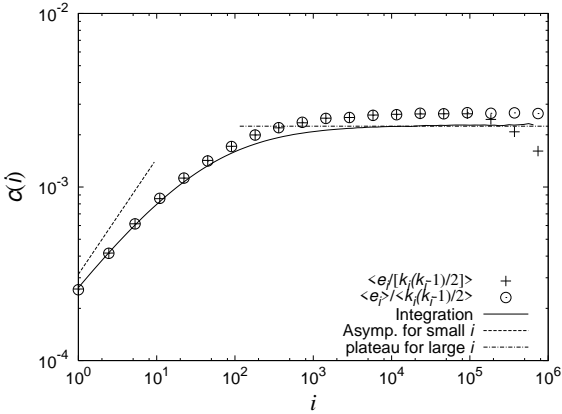}}
\caption{Plot of $C_i$. We can see that the approximation,
$\langle \frac{2e_i}{k_i(k_i-1)} \rangle \approx \frac{2\langle
e_i \rangle}{\langle k_i(k_i-1) \rangle}$ is valid for the large
$i$ limit. The dot-dashed line indicates the analytic result. In
the plateau region, the discrepancy between the analytic and the
numerical results decrease as system size $N$ increases. The
dashed line indicates the analytic results, Eqs.~(\ref{c-ii}) and
(\ref{eqq35}).} \label{fig4}
\end{figure}
\begin{figure}
\resizebox{0.9\columnwidth}{!}{\includegraphics{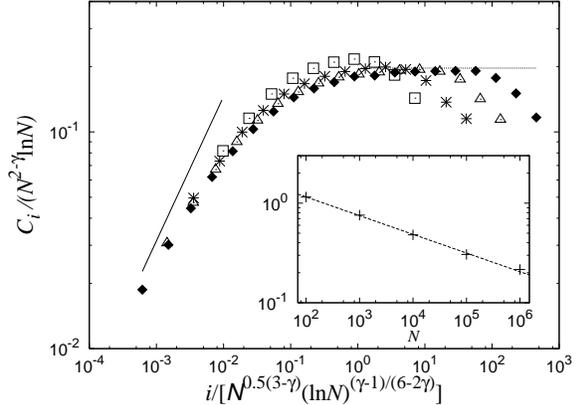}}
\caption{Size-dependence of local clustering coefficient $C_i$.
Data of various network sizes $N=10^3(\square)$, $10^4(\ast)$,
$10^5(\vartriangle)$ and $10^6(\blacklozenge)$ are collapsed in
the rescaling plot. Inset: Plot of the difference between the
analytic solution within the leading order and the simulation
value for $i=1$, divided by the simulation value as a function of
$N$. The decreasing behavior with increasing $N$ indicates that
the analytic solution is asymptotically valid.} \label{fig5}
\end{figure}
\begin{figure}
\resizebox{0.9\columnwidth}{!}{\includegraphics{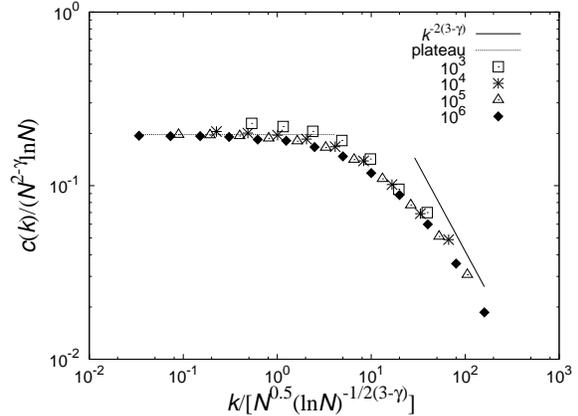}}
\caption{Plot of $C(k)$ for different system size
$N=10^3(\square)$, $10^4(\ast)$, $10^5(\vartriangle)$ and
$10^6(\blacklozenge)$. The data are well collapsed in the
rescaling plot.} \label{fig6}
\end{figure}

Next, the local clustering coefficient function $C_i$ is measured.
We first check the approximation introduced in
Eq.(\ref{c_i_approx}) in Fig.\ref{fig4}, finding that they overlap
each other except for large $i$. This discrepancy originates from
the fact that the vertices with large $i$ are mostly those located
at dangling ends with degree 1. Thus, the formation of triangles
or wedge shapes is rare and their numbers fluctuate highly. Next,
we also check the validity of the approximation from the discrete
summation and the continuous integration, Eq.(\ref{eqq20}). For
small $i$, the approximation is reasonably valid as shown in
Fig.\ref{fig4}, which can be expected in Appendix A. However, for
large $i$ in the flat region, the approximation shows some
discrepancy, but it is likely that the discrepancy decreases as
system size $N$ increases. To check the size-dependent behavior of
$C_i$, we plot $C_i$ versus $i$ with rescalings of $C_i
\rightarrow C_i/N^{1-\frac{1}{\mu}}\ln N$ and $i \rightarrow
i/\big(N^{(4\mu-4+\frac{1}{\mu})/(4\mu-2)}\ln^{1/(4\mu-2)} N\big)$
for different system sizes $N=10^3, 10^4$, $10^5$ and $10^6$. We
find that the data collapse reasonably well as shown in
Fig.\ref{fig5}. And we also check the behavior of $C(k)$. By
rescaling of $C(k)\rightarrow$ $C(k)/N^{1-\frac{1}{\mu}}\ln N$ and
$k\rightarrow k\ln^{1/2(2-\frac{1}{\mu})}/N^{1/2}$, the data of
$C(k)$ for different system sizes also collapse into a single
curve reasonably well as shown in Fig.\ref{fig6}. Thus, our
numerical simulation results show that, although several
approximations are involved in deriving the analytic results of
section 2, they are valid to the leading orders in $N$ as $N
\rightarrow \infty$.
\section{Conclusions and discussion}

We have studied analytically the mean neighboring degree function
$\bar k_{\rm{nn}}(k)$ and the clustering function $C(k)$ in the
static model for the case of $2 < \gamma < 3$ and checked the
results by numerical simulations. Due to the prevention of
self-loop and multiple edges, there occur intrinsic degree
correlations, which appear for $2 < \gamma < 3$ in the
$k$-dependent form of $\knn$ and $C(k)$ for large $k$. Our results
are summarized in Table I together with those for the case of
$\gamma \geq 3$. It would be interesting to compare our results
with those obtained in the generalized BA-type growth
model~\cite{barrat}. In this model, $\bar k_{\rm nn}(k)\sim
N^{(3-\gamma)/(\gamma-1)}k^{-(3-\gamma)}$ when $\gamma < 3$, $\sim
\ln N$ when $\gamma=3$, and $\sim \ln k$ when $\gamma
> 3$. On the other hand, $C(k)\sim
N^{(4-2\gamma)/(\gamma-1)}k^{-(3-\gamma)}$ for $k > (\ln
N)^{1/(3-\gamma)}$ and $\sim (\ln N)
N^{(4-2\gamma)/(\gamma-1)}k^{-2(3-\gamma)}$ for $k < (\ln
N)^{1/(3-\gamma)}$ when $\gamma < 3$, $\sim (\ln N)^2/N$ when
$\gamma=3$ and $\sim N^{-1}k^{\gamma-3}$ in the range $k <
N^{1/(\gamma-3)}$ when $\gamma > 3$. Therefore, it appears that
the degree correlation functions $\bar k_{\rm nn}(k)$ and $C(k)$
behave differently for the cases of the static model and the
BA-type growth model.

\renewcommand{\multirowsetup}{\centering}
\begin{center}
\begin{table*}
\caption{Degree and system-size dependence of $\bar k_{\rm
nn}(i)$, $\bar k_{\rm nn}(k)$, $C_i$ and $C(k)$.}
\begin{center}
\begin{tabular}{|c||c|c|c|c|}
    \multicolumn{5}{l}{$2<\gamma<3$}\\

    \hline \raisebox{-0.2ex}{Range} &
    \raisebox{-0.2ex}{$\overline{k}_{\textrm{nn}}(i)$} &
    \raisebox{-0.2ex}{$\overline{k}_{ \textrm{nn}}(k)$}
    &\raisebox{-0.2ex}{$C_{i}$}  & \raisebox{-0.2ex}{$ C(k)$}\\

    \hline \raisebox{-0.6ex}{$i>N^{3-\gamma}$} &
    \multicolumn{2}{c|}{\raisebox{-2.5ex}[0cm][0cm]{$\sim
    N^{\frac{3-\gamma}{\gamma-1}}$}}  &
    \multicolumn{2}{c|}{\raisebox{-2.5ex}[0cm][0cm]{$\sim N^{-(\gamma-2)} \ln N$}} \\

    $k<N^{\frac{\gamma-2}{\gamma-1}}$ & \multicolumn{2}{c|}{} &
    \multicolumn{2}{c|}{} \\

    \hline \raisebox{-0.3ex}{$N^{3-\gamma} > i > N^{\frac{1}{2}(3-\gamma)}$}
    & \multirow{4}{20mm}[-1mm]{$\sim N^{\frac{(3-\gamma)(\gamma-2)}{\gamma-1}}
    i^{\frac{3-\gamma}{\gamma-1}}$}
    & \multirow{4}{20mm}[-1mm]{$\sim N^{3-\gamma} k^{-(3-\gamma)}$}
    & \multirow{2}{35mm}[-1mm]{$\sim N^{-(\gamma-2)} \ln \left(\frac{
    i^{1/(\gamma-1)}}{N^{(3-\gamma)/(2\gamma-2)}} \right)$}
    & \multirow{2}{35mm}[-1mm]{$\sim N^{-(\gamma-2)} \ln \left(\frac{N^{1/2}}{k} \right)$} \\

    \raisebox{-0.3ex}{$N^{\frac{\gamma-2}{\gamma-1}}<k<N^{\frac{1}{2}}$} & & & &

    \\ \cline{1-1} \cline{4-5}
    \raisebox{-0.8ex}{$N^{\frac{1}{2}(3-\gamma)} >i$} &  &
    & \multirow{2}{35mm}[-1mm]{$\sim
    N^{-\frac{2\gamma^2+9\gamma+11}{\gamma-1}}i^{\frac{2(3-\gamma)}{\gamma-1}}$}
    & \multirow{2}{35mm}[-1mm]{$\sim N^{5-2\gamma} k^{-2(3-\gamma)}$} \\

    \raisebox{-0.6ex}{$N^{\frac{1}{2}}<k$} & & & & \\ \hline

    \multicolumn{5}{l}{$\gamma=3$}\\

    \hline \raisebox{-0.2ex}{whole range} & \multicolumn{2}{c|}{\raisebox{-0.2ex}{$\sim \ln N$}} &
    \multicolumn{2}{c|}{\raisebox{-0.2ex}{$\sim (\ln N)^2 /N$}} \\ \hline

    \multicolumn{5}{l}{$\gamma>3$}\\

    \hline \raisebox{-0.2ex}{whole range} & \multicolumn{2}{c|}{\raisebox{-0.2ex}{$\sim O(1)$}} &
    \multicolumn{2}{c|}{\raisebox{-0.2ex}{$\sim 1/N$}} \\ \hline
\end{tabular}
\end{center}
\end{table*}
\end{center}

\begin{acknowledgement} This work is supported by the KRF Grant No.
R14-2002-059-010000-0 in the ABRL program funded by the Korean
government MOEHRD.
\end{acknowledgement}

\appendix
\section{Transformation from discrete summation to continuous integration}
In several parts of this paper, we use the transformation from the
discrete summation to the continuous integration such as
\begin{equation}
    \sum_{j,k=1}^{N}F(i,j,k) \approx \int_{1}^{N}dj\int_{1}^{N}dk
    ~F(i,j,k).
    \label{appendix1}
\end{equation}
Here we discuss its validity. For a monotone decreasing function
$F(x)$, one has the well known relation:
\begin{equation}
\int_{1}^{N}dxF(x) + F(N) \leq \sum_{n=1}^{N}F(n) \leq
\int_{1}^{N}dx F(x) + F(1).
\label{appendix2}
\end{equation}
When $F(i,j,k)$ is positive, monotonously decreasing and bounded
in both $j$ and $k$, we can apply Eq.(\ref{appendix2}) twice to
obtain the error in Eq.(\ref{appendix1}) as
\begin{eqnarray}
&&\sum_{k}F(i,N,k)+\sum_{j}F(i,j,N) -F(i,N,N)\nonumber \\
 &&\leq \sum_{j,k} F(i,j,k) - \int_{1}^{N}dj
 \int_{1}^{N}dk F(i,j,k)\nonumber \\
&&\leq \sum_{k}F(i,1,k) +\sum_{j}F(i,j,1)-F(i,1,1).
\label{appendix3}
\end{eqnarray}
Thus Eq.(\ref{appendix1}) is valid when the ``surface terms'' in
Eq.(\ref{appendix3}) are negligible compared with the ``bulk
term'', $\int_1^N dj \int_1^N dk F(i,j,k)$. When we consider $\bar
k_{\rm nn}(i)$, $F(i,j,k)$ is given as $f_{ij}f_{jk}$ and one of
the surface terms that require special attention is
\begin{equation}
\sum_{k} F(i,1,k)=\sum_{k}f_{i1}f_{1k} \approx \frac{a^4
N^{3\mu-1}}{1-\mu}i^{-\mu}.
\end{equation}
with $a=\sqrt{2K(1-\mu)^2}$. It turns out that this surface term
is of the same order as the bulk term when $i > a^{\frac{2}{\mu}}
N^{2-\frac{1}{\mu}}$. Other surface terms are, however,
negligible. Thus, the contribution of the surface term to $\langle
k_{\rm nn} \rangle (i)$ is $\sim a^2 N^{2\mu-1}$. Then
Eq.~(\ref{eqq16-1}) for the case $i>
a^{\frac{2}{\mu}}N^{2-\frac{1}{\mu}}$ has to be changed as
\begin{equation}
    \frac{\int_{1}^{N} dj \int_{1}^{N}dk f_{ij}f_{jk}}{\langle k_{i}\rangle}
    \leq {\bar k_{\rm{nn}}} (i)
    \leq \frac{\int_{1}^{N}dj \int_{1}^{N}dk f_{ij}f_{jk}}{\langle k_{i}\rangle}
    +\frac{\sum_{k} f(i,1,k)}{\langle k_{i}\rangle}.
\end{equation}
This leads to
\begin{eqnarray}
    \frac{a^{2} N^{2\mu-1}}{2\mu-1}  & \leq {\bar k_{\rm{nn}}}(i) & \leq
    \frac{a^2 N^{2\mu-1}}{2\mu-1} \cdot 2\mu
    \label{appendix5}
\end{eqnarray}
to the leading order in $N$. Thus, the leading order of $\langle
k_{nn}\rangle (i)$ is given only in the form of the bounds when $i
> a^{\frac{2}{\mu}} N^{2-\frac{1}{\mu}}$.

When $C_i$ is considered, $F(i,j,k)=f_{ij}f_{jk}f_{ki}$. The most
relevant terms are $\sum_{k}F(i,1,k)$ and $ \sum_{j}F(i,j,1)$:
\begin{eqnarray}
&&\sum_{k}F(i,1,k)= \sum_{j}F(i,j,1)~~~~~~~~~~~~~~~~~~~~~~~~~~~~~~~~\nonumber \\
&&={}\sum_{k}f_{i1}f_{1k}f_{ki} \approx -q_0(\gamma)(\gamma
-1){a^{\frac{2}{\mu}}}N^{2-\frac{1}{\mu}}x^2/\mu,
\end{eqnarray}
when $i> a^{\frac{2}{\mu}} N^{2-\frac{1}{\mu}}$. These are of the
same order of magnitude as the bulk term in Eq.(\ref{eqq30}) up to
the $\ln N$ factor. Other boundary terms are smaller in order of
magnitude compared with these terms. Thus when $i>
a^{\frac{2}{\mu}} N^{2-\frac{1}{\mu}}$, $C_i$ is bounded as
\begin{eqnarray}{}
&&-q_0(\gamma)a^{\frac{2}{\mu}-2}(\gamma-1)(\gamma-2)^2
N^{1-\frac{1}{\mu}}
\ln(a^2 N^{2\mu-1}) \leq C_i~~~~~~~~~~~~~~~~~~ \nonumber \\
&&\leq -q_0(\gamma)a^{\frac{2}{\mu}-2}(\gamma-1)(\gamma-2)^2
N^{1-\frac{1}{\mu}}\left[ \ln(a^2 N^{2\mu-1}) +4 \mu \right].
\end{eqnarray}
The boundary term is important when $\ln N$ is not large enough
compared with $4\mu$.



\begin{thebibliography}{99}
\bibitem{rmp} R. Albert and A.-L. Barab\'asi, Rev. Mod. Phys.{\bf 74},
47(2002). 
\bibitem{porto} S. N. Dorogovtsev and J. F. F. Mendes, {\em Evolution of Networks} (Oxford University Press, Oxford,
2003). 
\bibitem{siam} M. E. J. Newman, SIAM Rev. {\bf 45}, 167 (2003). 
\bibitem{Barabasi99} A. -L. Barab\'asi and R. Albert, Science \textbf{286},
509 (1999); A. -L. Barab\'asi, R. Albert, and H. Jeong, Physica A
\textbf{272}, 173 (1999). 
\bibitem{vespignani} R. Pastor-Satorras, A. Vazquez and A. Vespignani,
Phys. Rev. Lett. {\bf 87}, 258701 (2001). 
\bibitem{Ravasz02a} E. Ravasz, A. L. Somera, D. A. Mongru, Z. N. Oltvai, and A. -L. Barab\'asi, Science \textbf{297}, 1551
(2002). 
\bibitem{Vazquez02} A. Vazquez, R. Pastor-Satorras and A.
Vespignani, Phys. Rev. E {\bf 65}, 066130 (2002).
\bibitem{Erdos59} P. Erd\H{o}s and A. R\'enyi,
Publicationes Mathematicae \textbf{6}, 290 (1959); Publications
of the Mathematical Inst. of the Hungarian Acad. of Sciences \textbf{5}, 17
(1960).
\bibitem{Goh01} K. -I. Goh, B. Kahng, and D. Kim, Phys. Rev. Lett.
\textbf{87}, 278701 (2001).
\bibitem{rome} G. Caldarelli, A. Capocci, P. De Los Rios, and M.A. Mu\~noz,
Phys. Rev. Lett. {\bf 89,} 258702 (2002).  
\bibitem{Soderberg02}B. S\H{o}derberg, Phys. Rev. E \textbf{66}, 066121 (2002). 
\bibitem{Boguna03} M. Bogu\~n\'a and R. Pastor-Satorras, Phys. Rev.
E \textbf{68}, 036112 (2003). 
\bibitem{nucl} D.-S. Lee, K.-I. Goh, B. Kahng and D. Kim,
Nucl. Phys. B {\bf 696}, 351 (2004). 
\bibitem{Maslov02} S. Maslov and K. Sneppen, Science \textbf{296}, 910 (2002). 
\bibitem{Newman02} M. E. J. Newman, Phys. Rev. Lett. \textbf{80}, 208701 (2002). 
\bibitem{Rowasz02} E. Ravasz and A.-L. Barab\'asi, Phys. Rev. E
\textbf{67}, 026112 (2003). 
\bibitem{Milo02} R. Milo et al., Science \textbf{298}, 824(2002). 
\bibitem{mich} M. Catanzaro and R. Pastor-Satorras,
Eur. Phys. J. B {\bf 44}, 241 (2005). 
\bibitem{Park03} J. Park and M. E. J. Newman, Phys. Rev. E {\bf
68}, 026112 (2003). 
\bibitem{barrat} A. Barrat and R. Pastor-Satorras, Phys. Rev. E
\textbf{71}, 036127 (2005).
\bibitem{szabo} G. Szab\'o, M. Alava, and J. Kert\'esz, Phys. Rev. E {\bf 67,}
056102 (2003).
\bibitem{boguna} M. Bogu\~n\'a, R. Pastor-Satorras, and A. Vespignani,
Eur. Phys. J. B {\bf 38,} 205 (2004). 
\bibitem{uncorr} M. Catanzaro, M. Bogi\~n\'a and R. Pastor-Satorras,
Phys. Rev. E \textbf{71}, 027103 (2005). 
\bibitem{MR} M. Molloy and B. Reed, Random Structure and Algorithms {\bf 6}, 161 (1995).
\bibitem{Ching} F. Chung and L. Lu, Annals of Combinatorics \textbf{6}, 125 (2002).
\end{thebibliography}
\end{document}